\documentclass[a4paper]{VisionStyle}
\usepackage{epsfig}

\begin{document}

\title{Source Detection in Simulated XMM-Newton Observations}

\author{J.\,Rasmussen\inst{1} \and K.\,Pedersen\inst{1} \and 
M.\,G\"{o}tz\inst{2}} 

\institute{
  Astronomical Observatory, University of Copenhagen, Juliane Maries Vej 30,
DK-2100 Copenhagen \O, Denmark
\and 
  Theoretical Astrophysics Center, Juliane Maries Vej 30,
DK-2100 Copenhagen \O, Denmark}

\maketitle 

\begin{abstract}

We present preliminary results from our on-going study: Comparing and
optimizing source detection procedures for XMM images.
By constructing realistic spatial and spectral source distributions
and ``observing'' these through the XMM Science Simulator we study how 
source characteristics and instrumental effects influence detection.
We are currently undertaking a statistical analysis on the outcome of running
source detection algorithms on the simulated EPIC pn and MOS images
in various energy bands. Particular emphasis is on the efficiency and 
reliability for detecting the faint, extended emission from distant clusters 
of galaxies.
Here we present simulated EPIC images applicable to the XMM Large-scale
Structure Survey (10 ksec pointings of ``blank fields'').
However, we emphasize that our procedure is flexible, including as many 
realistic 
source characteristics and instrumental effects as possible, and yet modest 
in computational demand. As such it can be used for simulating XMM data 
obtained from virtually any X-ray source.

All simulation products are made available at our website
({\tt http://www.astro.ku.dk/xcosmos/}).

\end{abstract}

\section{Introduction}
  
One of the main goals of the XMM Large-Scale Structure Survey (XMM LSS) 
(\cite{jrasmussen-WA3:pie01}) is to map the large-scale distribution of 
X-ray selected galaxy clusters out to a redshift of $\sim 1$ and to
detect the richest clusters out to a redshift of $\sim 2$.
Since the detection of faint, extended sources from X-ray data is a crucial
part of the XMM LSS project, an efficient and reliable source detection
method must be implemented for the survey. 

Previous investigations (\cite{jrasmussen-WA3:val01,jrasmussen-WA3:ref01}) 
have invoked simplified source field setups and left out 
some instrumental effects, when studying the efficiency of various detection
algorithms on simulated XMM images and estimating, e.g., completeness
and confusion limits.
By using schematic as well as more realistic source fields and
by a more complete treatment of instrumental effects through the 
XMM Science Simulator software {\sc SciSim} (which mimics the performance of 
XMM by 
ray-tracing incoming photons through the mirror and detector systems), 
we extend this earlier work and address additional questions, like
\begin{itemize}
\item What is the optimum energy band for distant cluster detection in 
XMM LSS data?
\item How does cluster X-ray morphology and off-axis distance influence
detection and photometric reconstruction of source properties?
\item How does one deal effectively with, e.g., blending of extended and 
pointlike sources and removal of bright point sources?
\end{itemize}

Here we present a few simulated 10 ksec XMM EPIC images being used for 
comparing and developing source detection procedures for the XMM Large Scale 
Structure Survey. 

\section{The Simulation Setup}
\label{jrasmussen-WA3_sec:setup}

\subsection{Source Fluxes}
\label{jrasmussen-WA3_sec:flux}

Three types of objects are considered, AGN type 1 and 2 and groups/clusters.
Point source (i.e.\ AGN) fluxes and object types are based on the X-ray 
Background (XRB) synthesis model of 
\cite*{jrasmussen-WA3:gil01}, down to a limiting flux of 
$1.3\times 10^{-15}$ erg cm$^{-2}$ s$^{-1}$ (0.5--2 keV). This XRB model is 
consistent with the projected cumulative number density $N$ of sources as a 
function of X-ray flux $S$ (the logN-logS relation) as
derived from deep XMM (\cite{jrasmussen-WA3:has01}) and Chandra 
(\cite{jrasmussen-WA3:toz02}) observations.
For clusters and groups, fluxes are provided by the assumed mass-temperature
($M-T$; see below) and luminosity-temperature ($L_X-T$) relations down to  
$5\times 10^{-15}$ erg cm$^{-2}$ s$^{-1}$ (0.5--2 keV).

We have produced two sets of simulations, distinguished only by the assumed 
cluster properties. In the first set, clusters are drawn from standard 
Press-Schechter (P-S) realizations of halos in a
flat $\Lambda$CDM cosmology with $\Omega_m=0.3$,  $\sigma_8=0.94$ and 
$H_0=65$ km s$^{-1}$ Mpc$^{-1}$. They are extracted from 100 Monte Carlo
realizations of the redshift interval $z=0-2$ down to a mass floor of
$10^{13}$ M$_{\sun} h_{100}^{-1}$, assuming the $M-T$ relation of 
\cite*{jrasmussen-WA3:eke96} and a non-evolving (local) cluster 
$L_X-T$ relation of $L_X \propto T^3$ (\cite{jrasmussen-WA3:bor99}) with no 
intrinsic dispersion.
To incorporate the observed $L_X-T$ dispersion as well as the steepening of 
the $L_X-T$ relation at low $T$ ($\leq 1$ keV), cluster fluxes are 
``corrected'' for these effects by assuming 
$L_X \propto T^{\alpha \pm \sigma}$, where $\sigma$ is the
standard deviation of a Gaussian distribution centered at $\alpha$.
For $T>1$ keV, $\alpha=3$ and $\sigma=0.15$ is assumed 
(consistent with results from ROSAT cluster surveys, cf.\ 
\cite{jrasmussen-WA3:bor99}), while $\alpha =5$ and $\sigma=1.0$ is taken for
$T<1$ keV, based on galaxy group observations (\cite{jrasmussen-WA3:hel00};
\cite{jrasmussen-WA3:xue00}). The cluster flux limit of 
$5\times 10^{-15}$ erg cm$^{-2}$ s$^{-1}$ is imposed hereafter.

The second set of simulations is based on a cosmological N-body simulation,
from which five fields covering $2.9 \times 2.9$ deg$^2$ and the redshift
interval $z \leq 1.4$ were extracted. Except for $\sigma_8 = 1.0$, other 
parameters are as above. Clusters are identified among the
$\sim 1.3\times 10^6$ dark matter particles in each field via a 
friends-of-friends algorithm (where gravitationally unbound particles are 
removed) and drawn from a random $30\arcmin \times 30\arcmin$ sight line 
through one of the fields.
The $L_X-T$ relation, including scatter as above, is here assumed 
to evolve such as to mimic no 
evolution in an ($\Omega_m=1,\Omega_{\Lambda}=0$) universe, i.e.\
$L_X \propto T^3 (d_{L} / d_L^{\star}) ^2$, where $d_L$ is the luminosity 
distance in an $(\Omega_m=0.3, \Omega_{\Lambda}=0.7)$ universe and 
$d_L^{\star}$ is the 
corresponding value in an $(\Omega_m=1,\Omega_{\Lambda}=0.0)$ cosmology.

A low-redshift cut at $z=0.1$ is imposed for computational reasons, 
excluding $\sim 2$\% and $<1$\% of the input clusters from the P-S and N-body 
calculations, respectively (i.e.\ in most LSS fields, such nearby 
groups/clusters will not appear).
 
\subsection{Spatial Flux Distributions}
AGN are assumed to be pointlike, whereas cluster surface brightness 
distributions follow a 2-D $\beta$--profile of core radius $r_c$.
Since there is to some extent a  'degeneracy' in the observational
determination of $\beta$ and $r_c$, we choose to fix $r_c$ and let only 
$\beta$ vary. For each cluster, values of $\beta$ and the ratio $\eta$ 
between minor and major isophote axis are drawn from observed 
distributions (\cite{jrasmussen-WA3:moh95}), assuming these to be Gaussians 
of mean 0.65 and $\sigma=0.16$ ($\beta$), and mean 0.80, $\sigma=0.12$ 
($\eta$). $r_c$ is fixed at 1/4 of the cluster radius, 
which is determined through the angular diameter distance relation 
assuming X-ray emission out to $r_{500}$. The latter is directly 
provided by the simulations in the N-body case, while it is calculated for 
the P-S clusters in the point mass approximation.

Groups and clusters are each represented by an ensemble of point sources 
spaced $2\arcsec$ apart, i.e.\ well below the on-axis PSF of FWHM 
$\simeq 6\arcsec$ 
(the in-flight measured mean value for the three X-ray telescopes at $E=1.5$ 
keV; \cite{jrasmussen-WA3:jan01}).

\subsection{Source and Background Spectra}
All sources are assigned a spectrum, with normalizations provided by the XRB 
model for point sources and by the P-S/N-body
simulations for clusters. The latter are assigned an {\sc xspec}-based 
{\em mekal} 
spectrum of metallicity $Z=0.25Z_{\sun}$, with temperature and redshift 
provided by the cosmological simulations.
For the AGN, a power law spectrum is assumed, i.e.\ $I(E)\propto E^{-\Gamma}$.
In consistency with observations 
(e.g., \cite{jrasmussen-WA3:pag98}), we choose $\Gamma=1.7$ for AGN1, 
while AGN2 are assigned a value $\Gamma=1.8$ and an intrinsic 
H~{\sc i} column density of 10$^{22}$ cm$^{-2}$. Given 
the spectral normalization in one band of the XRB model (e.g.\ 0.5-2 keV) 
this choice of spectra approximately reproduces the point source density 
predicted in the other model bands (2-10 and 5-10 keV).

A uniform diffuse X-ray background is added, modeled as a 
sum of two power laws. From fits to XRB observations we choose $\Gamma=1.4$ 
and a normalization $A$ at $E=1$ keV of 9.0 keV cm$^{-2}$ s$^{-1}$ sr$^{-1}$ 
keV$^{-1}$ for the power law dominating the XRB spectrum above $E\sim 1$ keV.
This effectively acts in the simulations as the integrated contribution of 
sources fainter than our adopted flux limit. A second power law
of $\Gamma=3.0$ and $A=1.0$ is also included, to account for the observed 
excess of emission at low 
energies (e.g., \cite{jrasmussen-WA3:miy98}; \cite{jrasmussen-WA3:par99})
compared to that expected from the $\Gamma=1.4$  power law. Adding this
component also
ensures an overall 1 keV normalization of 10.0 keV cm$^{-2}$ s$^{-1}$ 
sr$^{-1}$ keV$^{-1}$, consistent with most XRB measurements (e.g., 
\cite{jrasmussen-WA3:che97}). 

Both individual sources and the X-ray background are subjected to a uniform 
absorbing column density of $3\times 10^{20}$ 
cm$^{-2}$, which is the mean value expected for the XMM LSS 10 ksec pointings.
On top of this the in-orbit measured particle and internal 
background is added, using normalizations as given in the
XMM-Newton Users' Handbook\footnote{{\sf http://xmm.vilspa.esa.es/user/AO2/uhb/xmm\_uhb.html}} and spectra roughly modeled as power laws including Gaussian
representations of the four and two most prominent lines for the pn and MOS 
cameras, respectively (see also \cite{jrasmussen-WA3:kat02}; 
a more detailed background model is soon to be implemented). This 
background component is not subject to mirror vignetting and hence not 
included in the input to {\sc SciSim}.
 
\section{Data ``Reduction''}
Simulated data are created for the EPIC MOS and pn CCD's (with thin filters) 
using an exposure time of 10 ks, with all CCD's in full frame mode.
Spacecraft effects such as drift and jitter are neglected.
The output of {\sc SciSim} is, in accordance with real XMM data, in the form 
of Observation Data Files (ODF's) which include photon event files for each 
detector and CCD chip. Images and exposure maps are created from the event 
files using the XMM Science Analysis System ({\sc xmmsas}).

For illustration purposes a source field image is produced for each 
simulation. These are convolved with a $6\arcsec$ Gaussian and have image 
borders conforming to the field of view covered by the EPIC pn CCD's.

\section{Simulation Results and Source Detection}
\label{jrasmussen-WA3_sec:det}

\begin{figure}[!t]
  \begin{center}
    \epsfig{file=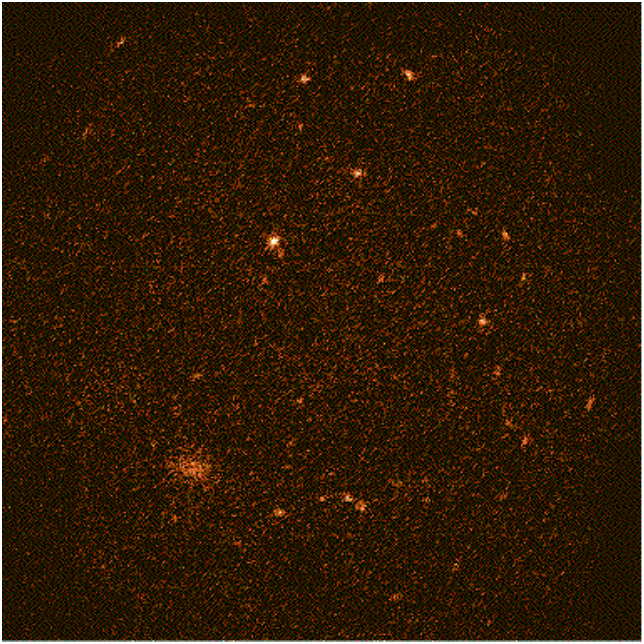, width=8.5cm}
  \end{center}
\caption{ Simulated 0.5--2 keV image (photon counts for pn+ 2MOS; logarithmic
 intensity scale). Spatial scale is $28\arcmin \times 28\arcmin$. }  
\label{jrasmussen-WA3_fig:fig1}
\end{figure}

\begin{figure}[!t]
  \begin{center}
    \epsfig{file=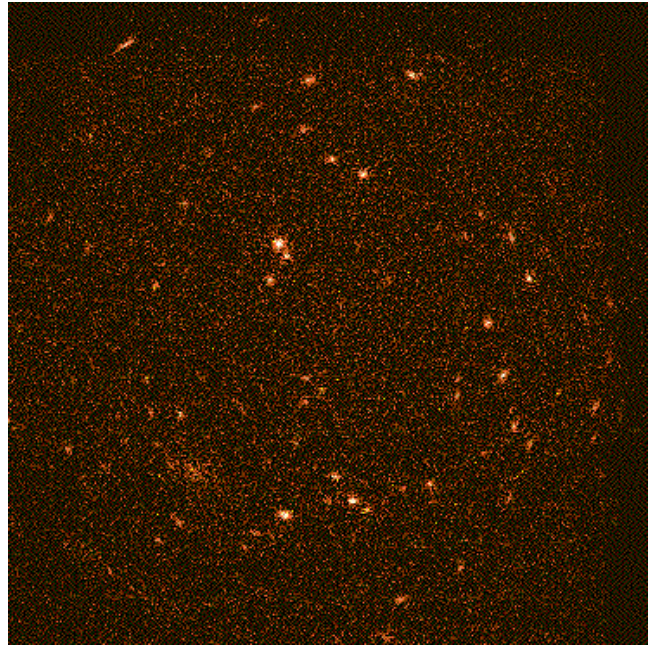, width=8.5cm}
  \end{center}
\caption{ {\em As Fig.~\ref{jrasmussen-WA3_fig:fig1}, but for the 2--10 keV 
band. }} 
\label{jrasmussen-WA3_fig:fig2}
\end{figure}

\begin{figure}[!t]
  \begin{center}
    \epsfig{file=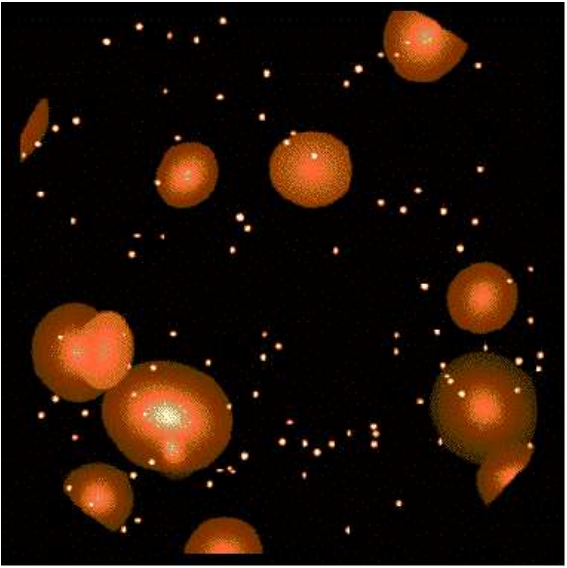, width=7.5cm}
  \end{center}
\caption{ Input source field, corresponding to the simulated images
shown in Figures~\ref{jrasmussen-WA3_fig:fig1} and 
\ref{jrasmussen-WA3_fig:fig2}. }   
\label{jrasmussen-WA3_fig:fig3}
\end{figure}

In Figures~\ref{jrasmussen-WA3_fig:fig1} and \ref{jrasmussen-WA3_fig:fig2}
we present examples of simulated EPIC images. These are ``raw'' 10 ksec 
photon count images, produced using the latest version (v3.0.0) of 
{\sc SciSim} with the XMM Current Calibration Files as of Dec 11, 2001, and
with the P-S prescription for extended sources 
(\S \ref{jrasmussen-WA3_sec:flux}).
For comparison, the corresponding input source field is shown in 
Figure~\ref{jrasmussen-WA3_fig:fig3}. In this particular representation, most
clusters are ``group-like'', with only two of the 12 input clusters having 
temperatures $T> 3$ keV. The brightest
one, a $T=3.3$ keV cluster at $z=0.35$ with a 0.5--2 keV flux of 
$8\times 10^{-14}$ ergs cm$^{-2}$ s$^{-1}$, stands out in the lower left 
corner.

\begin{figure}[!t]
  \begin{center}
    \epsfig{file=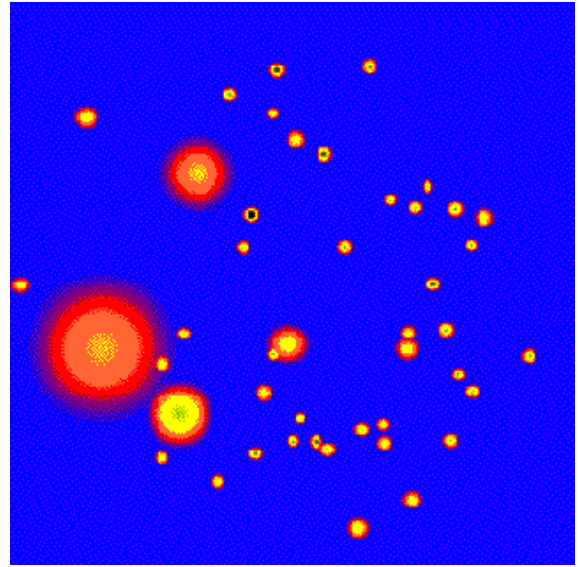, width=7.5cm}
  \end{center}
\caption{ Wavelet reconstruction of the image shown in 
Fig.~\ref{jrasmussen-WA3_fig:fig1}, illustrating the sources detected above
a $6\sigma$--significance. }   
\label{jrasmussen-WA3_fig:fig4}
\end{figure}

Based on the simulated images, we have done preliminary tests of the 
performance
of the wavelet--based {\sc xmmsas} source detection task {\em ewavelet},
using wavelet scales of 2, 4, 8, 16, and 32 pixels, and a detection 
significance threshold of $6\sigma$. 
This method was chosen as a starting point because it was demonstrated to 
work reasonably well for the simulated XMM images of 
\cite*{jrasmussen-WA3:val01}, and in this case, loosely judged, achieved 
optimum performance with the above 
parameters. A result of running this specific setup of {\em ewavelet} on the 
0.5--2 keV image of Fig.~\ref{jrasmussen-WA3_fig:fig1} is
illustrated in Figure~\ref{jrasmussen-WA3_fig:fig4}, showing the
wavelet-reconstructed image. Inspection of this image readily shows that
four extended sources have been ``detected''.
Comparing the result with the input source field 
in Fig.~\ref{jrasmussen-WA3_fig:fig3} reveals that one of these extended 
sources (the most central one) is actually two point sources blended together, 
whereas another results from the blending of two {\em extended} sources 
(this is not surprising, as one of these blended clusters is quite faint, 
being just above the adopted input flux limit of 
$5\times 10^{-15}$ ergs cm$^{-2}$ s$^{-1}$). However, the 
brightest cluster mentioned above is safely detected, and a relatively faint
($7\times 10^{-15}$ ergs cm$^{-2}$ s$^{-1}$),  $T=3$ keV cluster at 
$z=1.01$ is also found. The remaining extended sources are apparently either 
too faint, too far off-axis (clearly the case for at least five of them), or 
suffer too much blending with bright point sources to be detected in this 
case. We further note in passing that the number of point sources detected 
roughly corresponds to half the input number --- and this will certainly
improve when including the results based on the 2--10 keV image.

\section{Outlook}
\label{jrasmussen-WA3_sec:out}

The simulated 10 ksec EPIC images, examples of which have been shown in 
this paper, will provide a baseline for testing and 
optimizing cluster detection procedures for the XMM Large Scale Structure 
Survey. This will, among other things, allow us to firmly address the questions
posed in the introduction.

The detection of the $z\simeq 1.0$ cluster described above seems particularly 
promising with respect
to achieving our goal of finding high-redshift galaxy clusters, as it 
suggests that at least certain, reasonably massive but distant ($z>1$) clusters
could be well within the reach of detection in the XMM LSS Survey (and other
blank field XMM exposures of similar depth).
It should also be emphasized in this context that work still remains in order 
to optimize the 
method described here for the specific case of our simulated images. 

A comprehensive analysis is currently being carried out involving 
additional detection algorithms, and we are also testing how 
selected methods for image restoration (including the publicly available 
version of the {\sc pixon} code; {\tt http://www.pixon.com}) prior to source
detection may improve detection performance. Further plans for the near
future include testing source detection 
on deeper simulated exposures and to study how combining partially 
overlapping exposures will enhance detection. 
Results of this work will be presented in a forthcoming paper.

\begin{acknowledgements}

We thank R.\ Gilli for providing a numerical table of the XRB synthesis model
published in \cite*{jrasmussen-WA3:gil01}.
 
\end{acknowledgements}

\end{document}